\documentclass[11pt]{article}
\usepackage{amsmath}
\usepackage{hyperref}
\pdfoutput=1

\begin{document}
\title{Patterned coating by suspensions}
\author{Justin C.T. Kao and A. E. Hosoi\\
\\\vspace{6pt} Department of Mechanical Engineering
\\ Massachusetts Institute of Technology, Cambridge MA 02139, USA}
\maketitle

\begin{abstract}
  This file describes ``Patterned coating by suspensions'', a fluid
  dynamics video submitted to the Gallery of Fluid Motion of the 63rd
  Annual Meeting of the American Physical Society Division of Fluid
  Dynamics.
\end{abstract}

\thispagestyle{plain}
We examine the Landau-Levich flow of a suspension of neutrally buoyant
particles within a rotating glass cylinder. This differs from
previously investigated systems
\cite{denkov-et-al-1993:crystallization,buchanan-et-al-2007:pattern-draining}
in that we deal with macroscopic particles which immediately exceed
the thickness of the deposited liquid film. Thus, the presence of
particles on the liquid-coated wall creates excess surface area,
leading to a capillary attraction between particles. In turn, this
capillary attraction induces phase separation of the suspension
coating into regions of either dense particle clustering
or clear liquid. 

With the exceptions of the opening and closing credits, video in this
submission is of $\sim200 \mu\text{m}$ polystyrene beads suspended in
a mixture of water, sodium chloride, and Tween 20, with density $\rho
= 1.05\ \text{g}/\text{L}$. These sequences were filmed at 300 fps and
are played back 10$\times$ slower. The credits sequences are of $\sim
430 \mu\text{m}$ particles in a mixture of the same components, with
the addition of polyvinylpyrrolidone (Sigma PVP360) for increased
viscosity, recorded at 300 fps but played back in real time.

\enlargethispage*{2cm}

\end{document}